\documentclass[final,conference,twocolumn,letterpaper]{IEEEtran}

\IEEEoverridecommandlockouts              

\usepackage{fancyhdr}
\usepackage{graphicx}
\usepackage[utf8]{inputenc}

\graphicspath{{figures/}}

\usepackage{url}
\usepackage{wrapfig}
\usepackage{units}
\usepackage{caption}
\usepackage[printonlyused]{acronym}

\usepackage{calc}
\usepackage{graphicx}
\usepackage[lined,boxed, vlined,ruled,linesnumbered]{algorithm2e}
\usepackage{amsmath,amssymb,amsthm,amsfonts,amsbsy}
\usepackage{gensymb}

\usepackage{algpseudocode}
\usepackage{color}
\usepackage[table]{xcolor} 
\usepackage[caption=false]{subfig}
\usepackage{dblfloatfix}
\usepackage{nicefrac}
\usepackage{tcolorbox}
\usepackage{txfonts}

\usepackage{multirow}

\newcolumntype{C}[1]{>{\centering\arraybackslash}p{#1}}

\usepackage{booktabs}

\definecolor{BgGray}{gray}{0.7}%
\definecolor{BgGray2}{gray}{0.96}%
\definecolor{RowColorOdd}{named}{BgGray2}%
\definecolor{RowColorEven}{named}{white}%
\definecolor{comments}{gray}{.5}
\definecolor{Gray}{gray}{0.85}

\providecommand{\keywords}[1]{\textbf{\textit{Index terms---}} #1}

\usepackage{pifont}
\usepackage[multiuser,nomargin,marginclue,footnote]{fixme}
\fxusetheme{color}
\fxsetup{targetlayout=colorcb}
\FXRegisterAuthor{all}{anall}{ALL}
\FXRegisterAuthor{md}{anmd}{MD}
\FXRegisterAuthor{tolja}{antolja}{TOLJA}
\FXRegisterAuthor{pg}{anpg}{PG}
\FXRegisterAuthor{mch}{anmc}{MC}
\FXRegisterAuthor{cp}{ancp}{CP}
\FXRegisterAuthor{awo}{anawo}{AWO}

\fxsetup{status=draft}

\begin{document}
\providetoggle{techreport}
\settoggle{techreport}{false}

\title{On Phase Offsets of 802.11ac Commodity WiFi}

\author{
\IEEEauthorblockN{Anatolij Zubow, Piotr Gaw{\l}owicz, Falko Dressler}
\IEEEauthorblockA{Technische Universität Berlin, Germany}
\{zubow, gawlowicz, dressler\}@tkn.tu-berlin.de
}

\maketitle


\begin{abstract}
%
%
We analyze the phase offsets between RF chains of modern IEEE 802.11ac chips. 
We investigate both the 2.4 and 5\,GHz bands on a per OFDM subcarrier level.
Results reveal that the phase offset between receive antennas is due to random phase rotations semi-time-invariant with up to four possible values.
Moreover, it is frequency-dependent. 
We propose a simple algorithm, which allows us to correct the phase offset on the fly without any calibration.
As proof-of-concept, we implemented Angle of Arrival (AoA) using MUSIC algorithm.
To achieve higher accuracy we stitched the thirteen overlapping 20\,MHz channels available in 2.4\,GHz band together to effectively have a single 80\,MHz channel.
Results show very good AoA precision although only two receive antennas were used.
\end{abstract}

\keywords{WiFi, 802.11, visible light communications, LiFi, COTS, testbed}

%

\section{Introduction}
\begin{figure}[ht]
    \centering
    \includegraphics[width=0.85\linewidth]{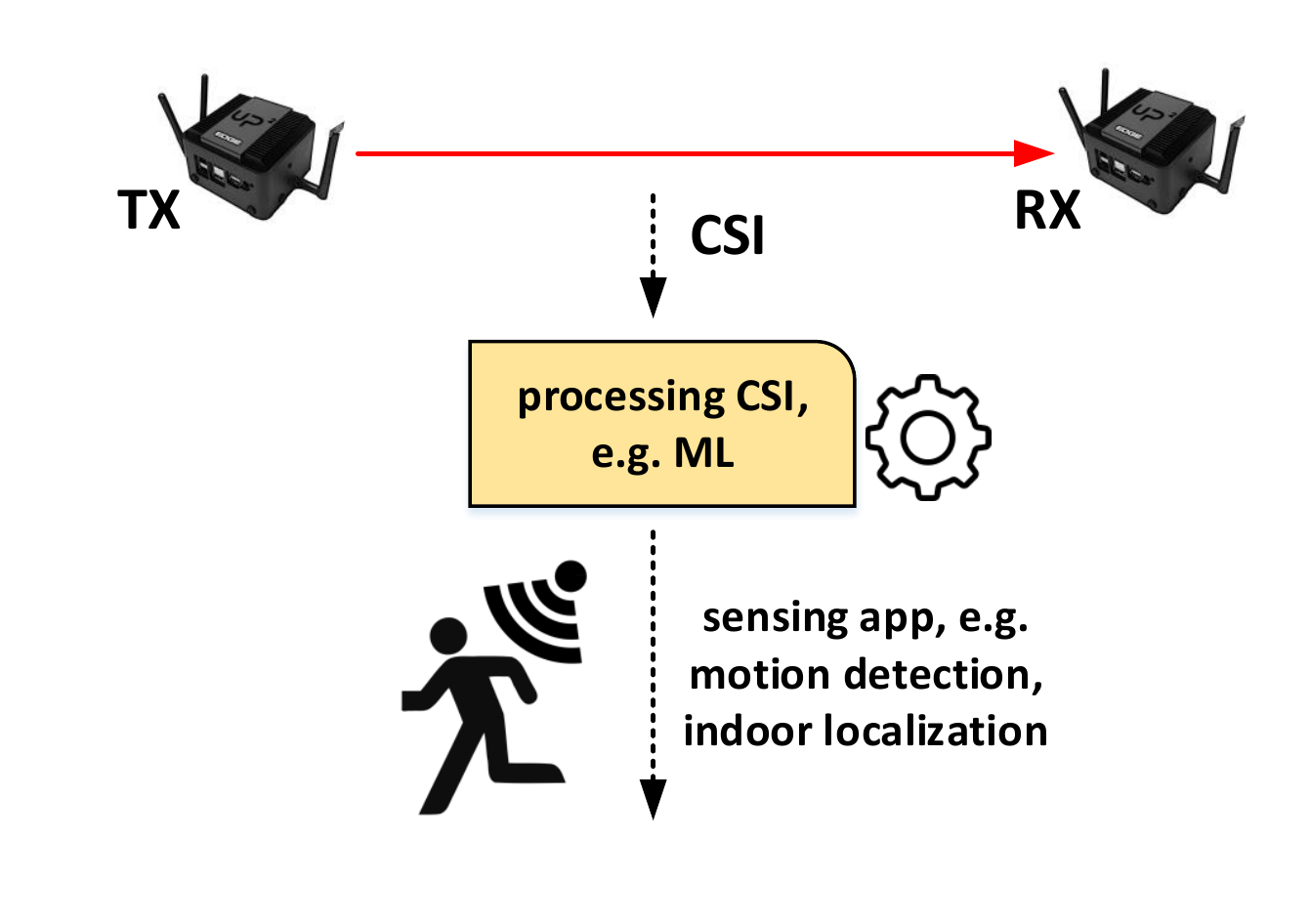}
    \caption{CSI processing in wireless sensing applications.}
    \label{fig:csi_apps}
    \vspace{-5pt}
\end{figure}

In recent years we have seen a boom of wireless sensing applications ranging from user localization and tracking, line-of-sight path identification, passive human sensing, motion recognition and wellness monitoring~\cite{ma2019wifi} (Fig.~\ref{fig:csi_apps}). 
An indoor localization system (ILS) based on existing and already deployed WiFi infrastructure would be very promising as indoor localization might become ubiquitous to any device equipped with a WiFi chip (e.g., smartphone, tablet) like the Global Positioning System (GPS), which is used outdoors.
However, such an ILS needs to be accurate, deployable and universal~\cite{kotaru2015spotfi}.
Recent localization techniques that rely on angle of arrival (AoA) estimation satisfy all the three requirements.
The AoA schemes utilize the Channel State Information (CSI) captured by the multiple antennas of commodity WiFi devices.

An important obstacle, which prevents many AoA approaches from practical deployment on commodity WiFi devices is the phase offset between RF chains in WiFi chips~\cite{zhang2019calibrating}.
This is because the signals received from different antennas are processed by different RF chains independently; the measured CSI will be distorted by the phase offsets between RF chains.
The focus of this paper is to analyze the difference of the initial phase offsets on different RF chains.
For old generation of 802.11n WiFi devices this was analyzed by Zhang et al.~\cite{zhang2019calibrating}.
They found out that in commodity 802.11n chips like Intel 5300 and Atheros AR9380 the phase offsets are semi-time-invariant with two possible values and hence semi-deterministic.
%

The scope of this paper is to analyze the phase offset between receive antennas of modern commodity 802.11ac chips like Intel 9260.
Therefore, we present results analyzing the RX phase offset on a per-OFDM subcarrier level and not just channel granularity as in~\cite{zhang2019calibrating}.
Our results reveal that the RX antenna phase offset of COTS 802.11ac Intel 9260 is semi-time-invariant as well but with up to four possible values which is different to the old 802.11n chips (Intel, Atheros) having just two possible values. 
Moreover, it depends on the frequency (i.e., RF channel/subcarrier) used.
Its semi-deterministic characteristic allows use to correct the phase offset on the fly without any calibration.
As proof-of-concept we implemented AoA using MUSIC algorithm in 2.4\,GHz band.
Therefore, we stitched together the CSI from all thirteen overlapping 20\,MHz channels from 2.4\,GHz band to effectively have a single 80\,MHz channel.
Results show very good AoA precision although only two receive antennas were used.

\medskip

\noindent \textbf{Contributions:}
First, we analyze the characteristics of RX phase offset of COTS 802.11ac using the Intel 9260 COTS chip.
Second, we present an algorithm for cleansing the CSI from random phase rotation introduced by the COTS chip to derive the true RX phase offset.
Third, using measurements we present the true phase offsets between the two RX antennas of COTS 802.11ac (Intel 9260) chip.
Forth, as proof-of-concept we implemented an algorithm for AoA estimation to show that CSI obtained from COTS 802.11ac and cleansed can be used to give precise AoA.

%

\section{Problem Statement}
An RF signal generated by the transmitter propagates through multiple paths, such as direct propagation (radiation), reflection and scattering, and superimposes at the receiver, carrying the information of the characteristics of the propagated environment, the so-called Channel State Information (CSI).
However, the CSI obtained from COTS WiFi chip characterizes not only the frequency response of the wireless channel, but also contains several kinds of phase distortion introduced by the imperfect inertial circuits~\cite{zhang2019calibrating}.
According to \cite{zhuo2017perceiving,zhang2019calibrating} the carrier frequency offset (CFO), packet detection delay (PDD), and sampling frequency offset (SFO) do not cause catastrophic problem to the AoA algorithms since they are the same among different RF chains.
However, the phase locked loop (PLL) initial phase is different among RF chains.
The focus of this paper is to analyze the characteristics of phase offsets among different RF chains due to PLL initial phase difference.
For old generation of COTS WiFi devices this was analyzed by Zhang et al.~\cite{zhang2019calibrating}.
They found out that in COTS 802.11n chips like Intel 5300 and Atheros AR9380 the phase offsets are semi-time-invariant with two possible values and hence semi-deterministic.
The goal of this paper is to make a similar study for modern 802.11ac COTS chips like the Intel 9260.
Moreover, we want to come up with solutions making the use of CSI suitable for the usage in AoA algorithms.

%

\section{Platform}\label{sec:platform}
\begin{figure}[ht]
    \centering
    \includegraphics[width=0.7\linewidth]{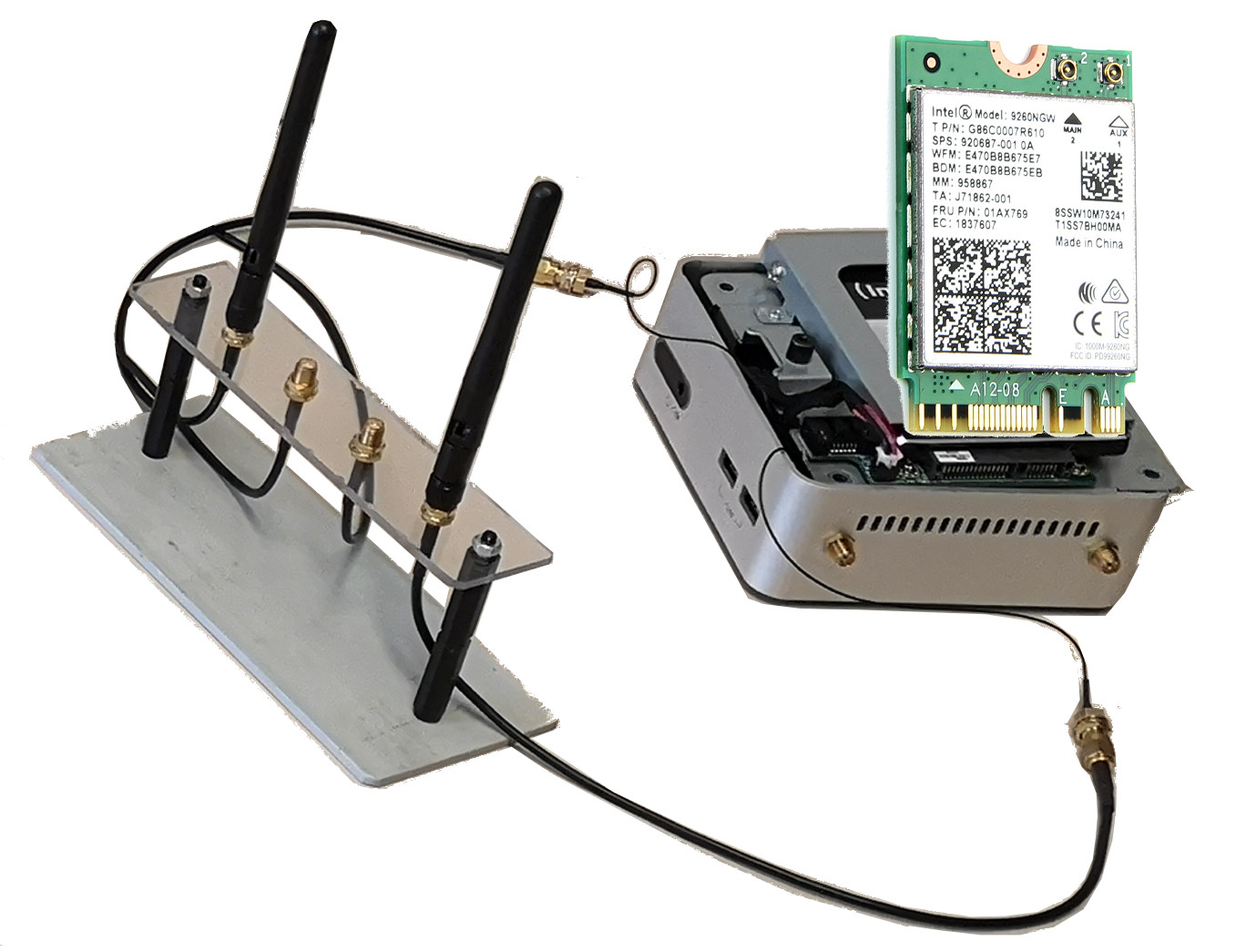}
    \caption{The experimental device with 802.11ac chip (Intel 9260) \& antenna array with two elements.}
    \label{fig:platform}
    \vspace{-5pt}
\end{figure}

As experimentation platform, we used mini computers (Intel NUC) equipped with Intel 9260 WiFi NICs (Fig.~\ref{fig:platform}).
The Intel 9260 is an IEEE 802.11ac wave 2 compliant radio with 2x2 MIMO, channel width of up-to 160\,MHz and support for multi-user MIMO.
A pair of such nodes was used during the experiments.
As the CSI functionality was not available for the Intel 9260 WiFi chip we had to port them from Intel backport drivers\footnote{https://git.kernel.org/pub/scm/linux/kernel/git/iwlwifi/backport-iwlwifi.git} release/core46 to the Linux 5.5.1 kernel.
The Ubuntu desktop 18.04 OS together with our patched Linux kernel was used for both the transmitter and the receiver.
We run both the transmitter and receiver in \texttt{monitor} mode.
For each received packet the CSI was estimated by the WiFi driver and passed to the Linux user space using Netlink API.
Here the Netlink messages were received and processed using the UniFlex~\cite{gawlowicz2017uniflex} control framework written in Python.
We had to reverse engineer the encoding of the CSI as the CSI message format was not provided by Intel.
To proof the correctness we conducted extensive measurements over cable/air setups and compared the results with the Linux 802.11n CSI Tool~\cite{halperin2011tool} using the old 802.11n Intel 5300 NIC.


%

\section{RX phase offset characteristics}\label{sec:results}
In order to understand the phase offset between receive chains (antennas) of IEEE 802.11ac COTS WiFi chips we conducted experiments.

\subsection{Experiment}
\begin{figure}[ht]
    \centering
    \includegraphics[width=0.8\linewidth]{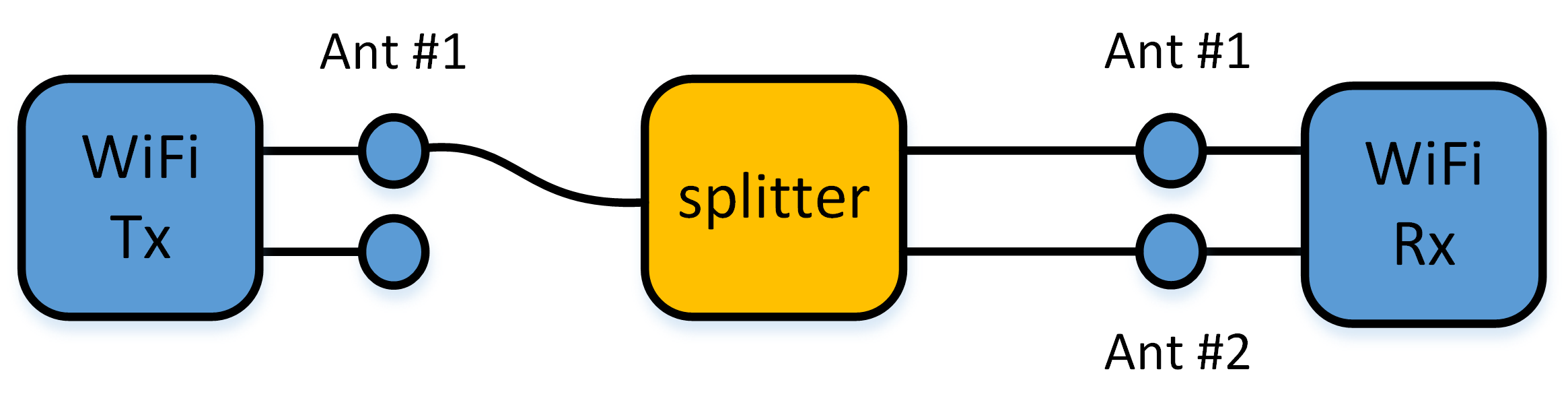}
    \caption{Experiment setup: transmitter connected via coaxial cables and splitter to receiver.}
    \label{fig:cable_exp_setup}
    \vspace{-5pt}
\end{figure}

We used a pair of nodes based on our platform (\S\ref{sec:platform}).
In order to avoid the influence caused by the environment (i.e., multipath propagation) and its changes (e.g., mobility), the receiver and transmitter are connected via coaxial cables and splitters (Fig.~\ref{fig:cable_exp_setup}).
On the transmitter side only the first antenna port was used.
With such a setup, the measured phase offset between receive antennas may also contain the constant phase offsets introduced by cables and splitters.
However, by using the same cables and splitters, such additional offsets remain the same during experiments, and will not affect the result.
For the elimination of these offsets we used the technique method from \cite{xiong2013arraytrack,zhang2019calibrating} which swaps the
external cables at the splitter and averages the measurement results.
%
%
We performed measurements on all WiFi channels available on our Intel 9260 WiFi NIC, namely:
\begin{itemize}
\item 2.4\,GHz band: channels 1-13,
\item 5\,GHz band: channels 36-64 and 100-165,
\end{itemize}
So in total 580\,MHz of spectrum were measured.
On each 20\,MHz channel 10k packets (HT20, MCS 0, BPSK 1/2) were send. 
After finishing transmitting packets the channel was switched by the transmitter locally and remotely on the receiver using the Uniflex framework. 
From each received packet we collected the CSI and annotated with the channel used.
For post-processing we used Matlab.

\subsection{Results}
\begin{figure}[ht]
    \centering
    \includegraphics[width=1\linewidth]{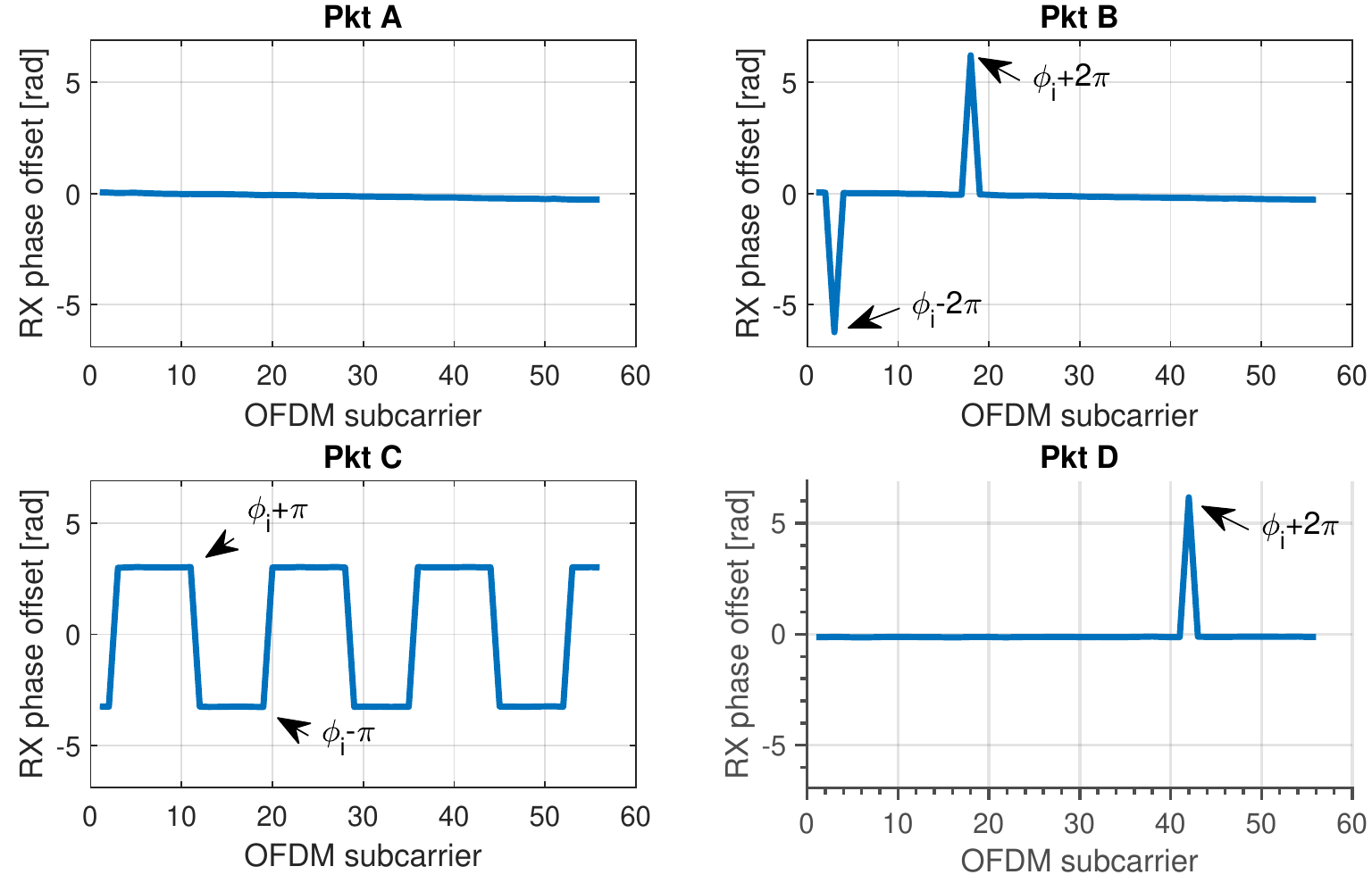}
    \vspace{-10pt}
    \caption{RX phase offset measured from four different packets.}
    \label{fig:show_phase_offset_rnd}
    \vspace{-5pt}
\end{figure}

The results from experiments show that the measured rx phase offset $\hat{\phi}$ is random but semi-deterministic.
The value of the true phase offset $\phi$ may rotate by multiple of $\pi$. 
Fig.~\ref{fig:show_phase_offset_rnd} shows the measured $\hat{\phi}$ of four arbitrarily selected packets for each OFDM subcarrier:
\begin{itemize}
\item \textbf{A:} the perfect case with $\hat{\phi} = \phi$, i.e. measured phase offset equals the true one,
\item \textbf{B:} all subcarrier are correct except that subcarriers 3 and 18 in $\hat{\phi}$ are phase offset rotated by $-2\pi$ and $+2\pi$ respectively,
\item \textbf{C:} the erroneous case with phase rotated on each subcarrier by $\pi$, i.e. not a single subcarrier in $\hat{\phi}$ has correct phase offset (cf. A),
\item \textbf{D:} similar to case A with a single subcarrier rotated by $+2\pi$.
\end{itemize}
From the experimental results we can conclude that the measured rx phase offset $\hat{\phi}$ may rotate by multiple $\pi$ around $\phi$.
Moreover, by analyzing the whole data set we discovered that the rotation strictly depends on whether the true $\phi$ is positive or negative:
\begin{align}
\hat{\phi} =
\begin{cases}
\phi + n\pi, n \in \{-2,-1,0,1\}, & \text{if  } \phi \geqslant 0 \label{eq:phase_eq}\\ 
\phi + n\pi, n \in \{-1,-1,0,1,2\}, & \text{otherwise}.
\end{cases}
\end{align}
where $\hat{\phi}$ and $\phi$ are the measured and the true rx phase offset respectively.
Our observation is similar to the one made by Zhang et al. for 802.11n chips~\cite{zhang2019calibrating} except that we also observe rotation my more than one $\pi$ (Fig.~\ref{fig:show_phase_offset_rnd}, packets B \& D).
So the measured $\hat{\phi}$ can have up to four possible values.

\begin{figure}[ht]
    \centering
    \includegraphics[width=1\linewidth]{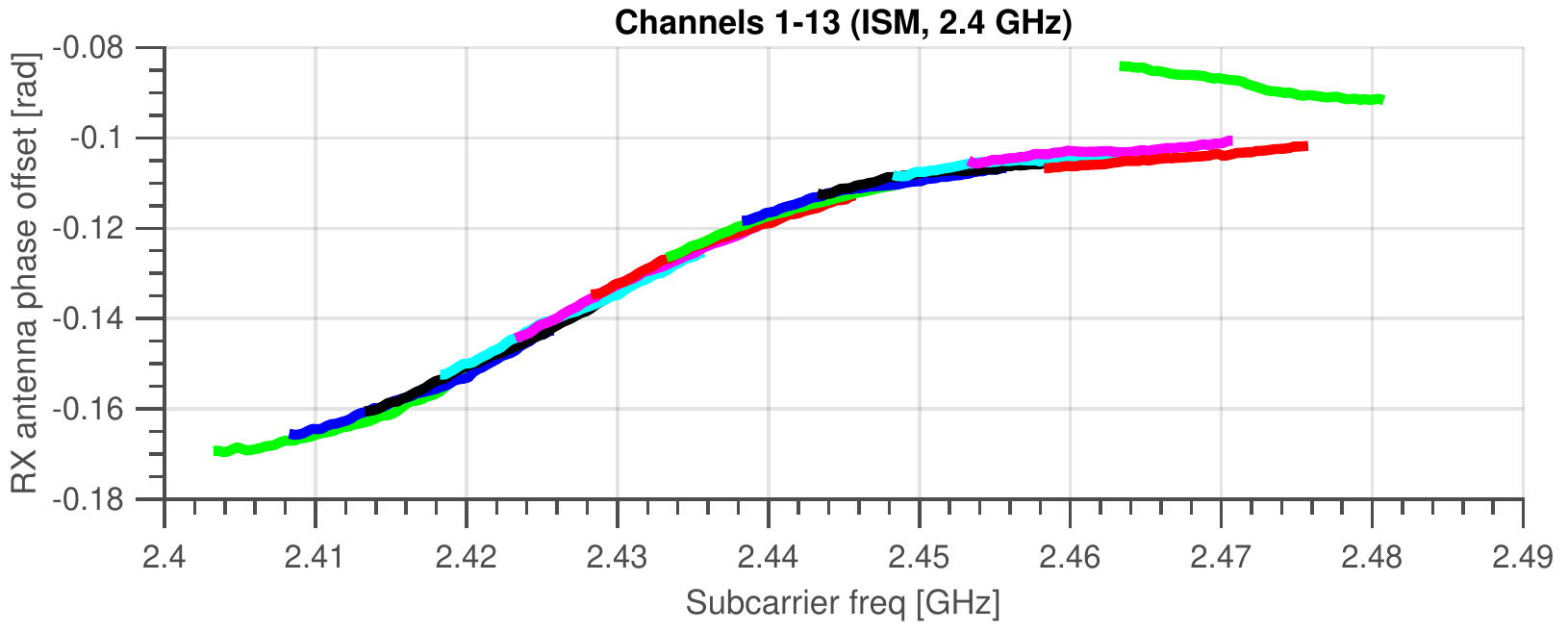}
    \vspace{-10pt}
    \caption{Phase offset between the two receive antennas in 2.4\,Ghz band.}
    \label{fig:rxphaseoffset_24ghz}
    \vspace{-5pt}
\end{figure}

In order to obtain the true rx phase offset, $\phi$, from our cable experiment we had to clean up the data from the random phase rotation.
A simple cleansing approach was possible here as we know the true $\phi$ to be around zero as we use cables.
Hence all samples with invalid values were discarded.
Fig.~\ref{fig:rxphaseoffset_24ghz} shows the median phase offset between the two receive antennas of the WiFi NIC for the channels 1-13 in 2.4\,GHz ISM band on a per subcarrier basis.
The figure shows the phase offset after elimination of phase offsets introduced by cables and splitters.
We can observe that the RX phase offset is close to zero and slightly frequency-dependent.
Tab.~\ref{fig:tbl24ghz} summarizes the results of measured values of phase offsets per channel by averaging over the subcarriers.
The difference in phase offset among different subcarriers in 2.4\,GHz band is small, i.e. $0.086 \approx 5$\degree. 
\begin{table}
\begin{tabular}{@{} l *{7}c @{}}
\toprule
 \multicolumn{1}{c}{Channel}    & 1  & 2  &  3 &  4  &  5 &  6 \\ 
\midrule
 $\phi$ (rad) & -0.1628  & -0.1557  &  -0.1477 &  -0.1388  &  -0.1303 &  -0.1228  \\ \bottomrule
 \end{tabular}
 \begin{tabular}{@{} l *{13}c @{}}
\toprule
 \multicolumn{1}{c}{Channel}    & 7 &  8 &  9 &  10 &  11 &  12 \\ 
\midrule
 $\phi$ (rad) & -0.1163 &  -0.1114 &  -0.1079 &  -0.1054 & -0.1031 &  -0.1044 \\ \bottomrule
 \end{tabular}
 \begin{tabular}{@{} l *{2}c @{}}
\toprule
 \multicolumn{1}{c}{Channel}    & 13 \\ 
\midrule
 $\phi$ (rad) & -0.0883 \\ \bottomrule
 \end{tabular}
    \caption{Receive antenna phase offset in 2.4\,Ghz band after elimination of randomness and phase offsets introduced by cables and splitters.}
    \label{fig:tbl24ghz} 
\end{table}

\begin{figure}[ht]
    \centering
    \includegraphics[width=1\linewidth]{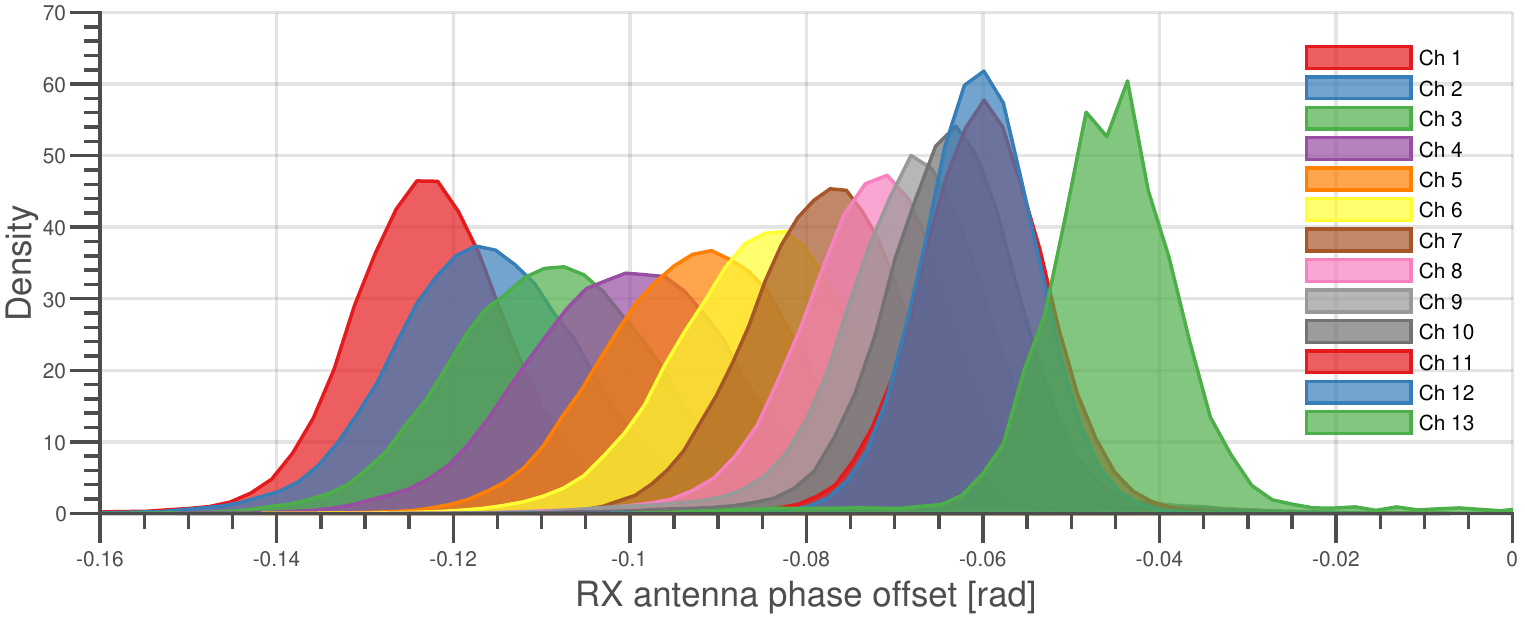}
    \vspace{-10pt}
    \caption{Distribution of RX antenna phase offset (2.4\,GHz).}
    \label{fig:rxoffset_density}
    \vspace{-5pt}
\end{figure}

Finally Fig.~\ref{fig:rxoffset_density} shows the distribution of the valid phase offsets, i.e. those without random phase offset rotation.
We see a narrow distribution, i.e. more data closer to the mean, with a standard deviation of just $0.0008 \approx 0.05$\degree.
%

\begin{figure}[ht]
    \centering
    \includegraphics[width=1\linewidth]{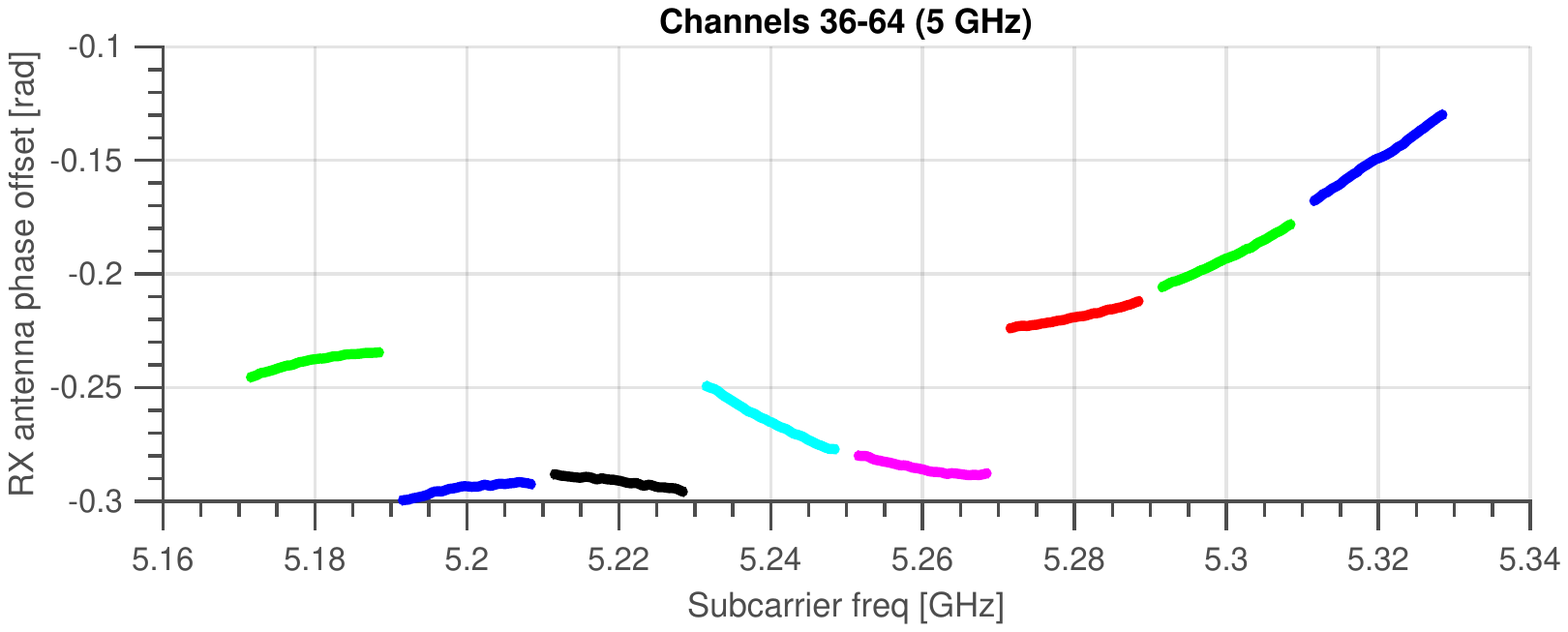}
    \vspace{-10pt}
    \caption{Phase offset between the two receive antennas in 5\,Ghz band (channels: 36-64).}
    \label{fig:rxphaseoffset_5aghz}
    \vspace{-5pt}
\end{figure}

\begin{figure}[ht]
    \centering
    \includegraphics[width=1\linewidth]{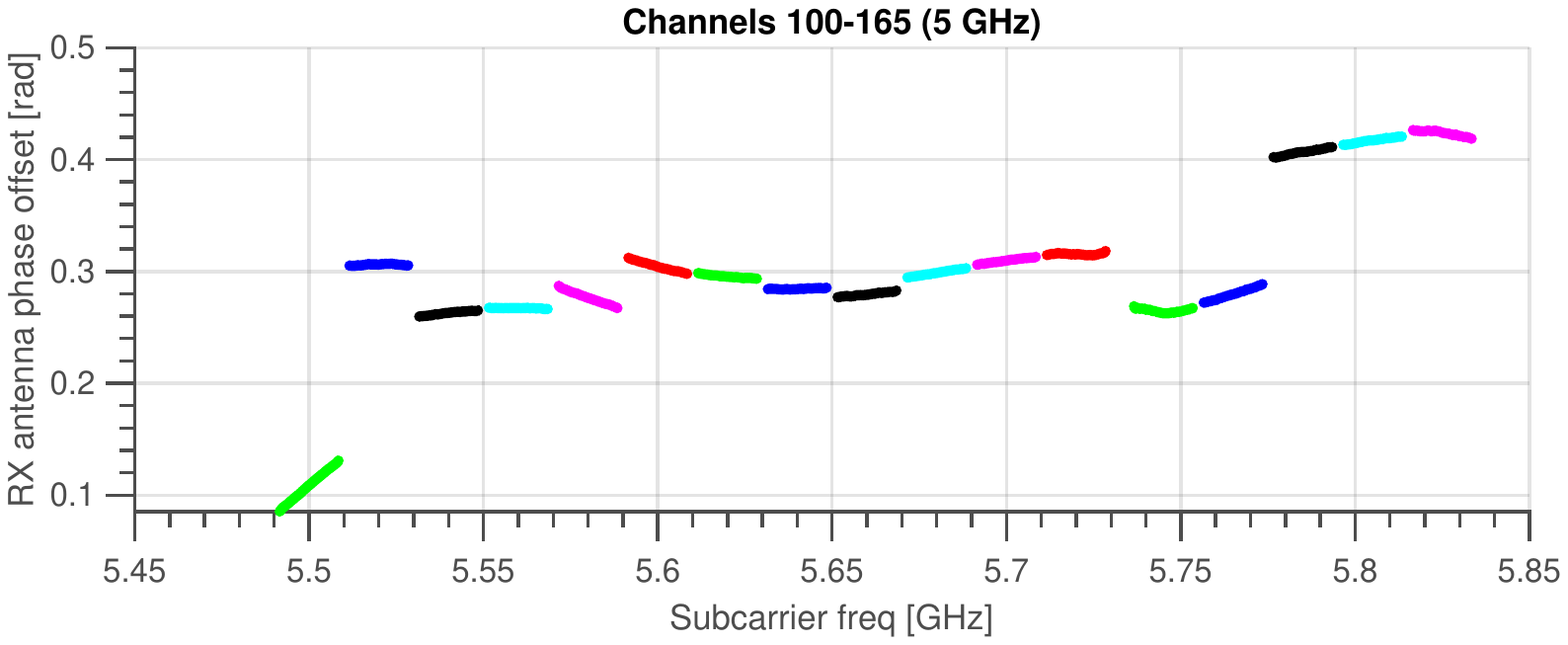}
    \vspace{-10pt}
    \caption{Phase offset between the two receive antennas in 5\,Ghz band (channels: 100-165).}
    \label{fig:rxphaseoffset_5bghz}
    \vspace{-5pt}
\end{figure}

Finally, Fig.~\ref{fig:rxphaseoffset_5aghz} and Fig.~\ref{fig:rxphaseoffset_5bghz} shows the values for the channels in 5\,GHz band.
We can see that the lower 5\,GHz channels have an rx phase offset between -0.3 and -0.1 the higher channels are between 0.1 and 0.5.
Note that such range equals $45\degree$.

%

\section{RX phase offset correction}\label{sec:cleansing}
\begin{figure}[ht]
    \centering
    \includegraphics[width=1\linewidth]{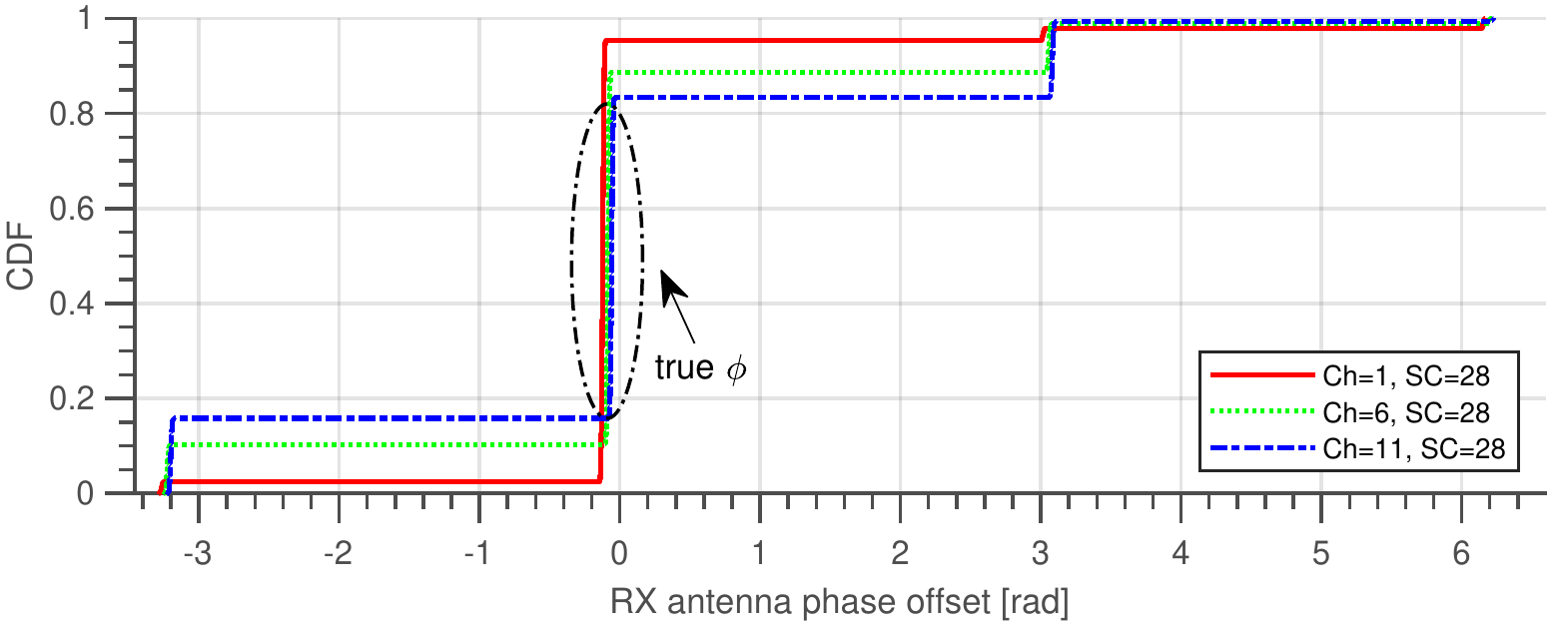}
    \vspace{-10pt}
    \caption{Distribution of $\hat{\phi}$ measured over 10k packets on subcarrier 28 for channels 1, 6 and 11.}
    \label{fig:show_phi_pdf}
    \vspace{-5pt}
\end{figure}

As the measured rx phase offset $\hat{\phi}$ experiences random phase rotation it needs to be corrected before it can be used by AoA algorithms.
Our proposed approach is based on the following key observations.
First, $\hat{\phi}$ is semi-time-invariant with four possible values and hence semi-deterministic.
%
Second, altough some subcarrier may be randomly rotated at some point in time, the measured phase offset is correct for the majority of time.
As an example Fig.~\ref{fig:show_phi_pdf} shows the distribution of $\hat{\phi}$ for the channels 1, 6 and 11 measured on subcarrier 28.
Here we see that the majority of $\hat{\phi}$ had the correct phase, i.e. $\hat{\phi} = \phi \approx 0$, and in only $<$30\% of the cases a wrong $\phi$ was reported.
Hence, my measuring $\hat{\phi}$ from sufficient large number of packets the effect of random phase rotation can be averaged out.
Specifically, we measure a particular OFDM subcarrier multiple times, i.e. sending multiple packets on same or overlapping channel and combine results using detection and replacement of outliers in data and merge into final value using \texttt{median} operator:
\begin{itemize}
\item In 2.4\,GHz band neighboring channels are overlapping by 15\,MHz, i.e. a particular frequency (OFDM subcarrier) is measured by multiple channels. We exploit that by sending packets on all channels 1 to 13.
\item To further compensate for random phase rotation we send $N$ packets on each channel.
\item We end up having the rx phase offset $\hat{\phi}_{t,c}(f)$ measured for a particular subcarrier multiple times where $f$ is OFDM subcarrier, $t=1 \ldots N$ and $c \in C(f)$ the set of channels being overlapping on that subcarrier.
\item We construct a vector $\overrightarrow{\hat{\phi}_{t,c}} = (\hat{\phi}_{t,c}(0), \ldots, \hat{\phi}_{t,c}(F))$ where $0 \ldots F$ represents the total amount of subcarriers when combining the 13 channels.
\item Next we filter out and replace the outliers $\overrightarrow{\phi_{t,c}^{*}} = \mathrm{filloutliers}(\overrightarrow{\hat{\phi}_{t,c}})$.
\item The rx phase offset for a particular subcarrier is estimated by computing the median $\phi^{*}(f) = \mathrm{med}(\phi_{t,c}^{*}(f)), t=1 \ldots N, c \in C(f)$.
\item The final vector representing the corrected phase offset is created as $\overrightarrow{\phi} = \mathrm{filloutliers}((\phi^*(0), \ldots, \phi^*(F)))$.
\end{itemize}
Note, for the filtering we used function \texttt{filloutliers(x,'linear')} from MatLab.

\begin{figure}[ht]
    \centering
    \includegraphics[width=1\linewidth]{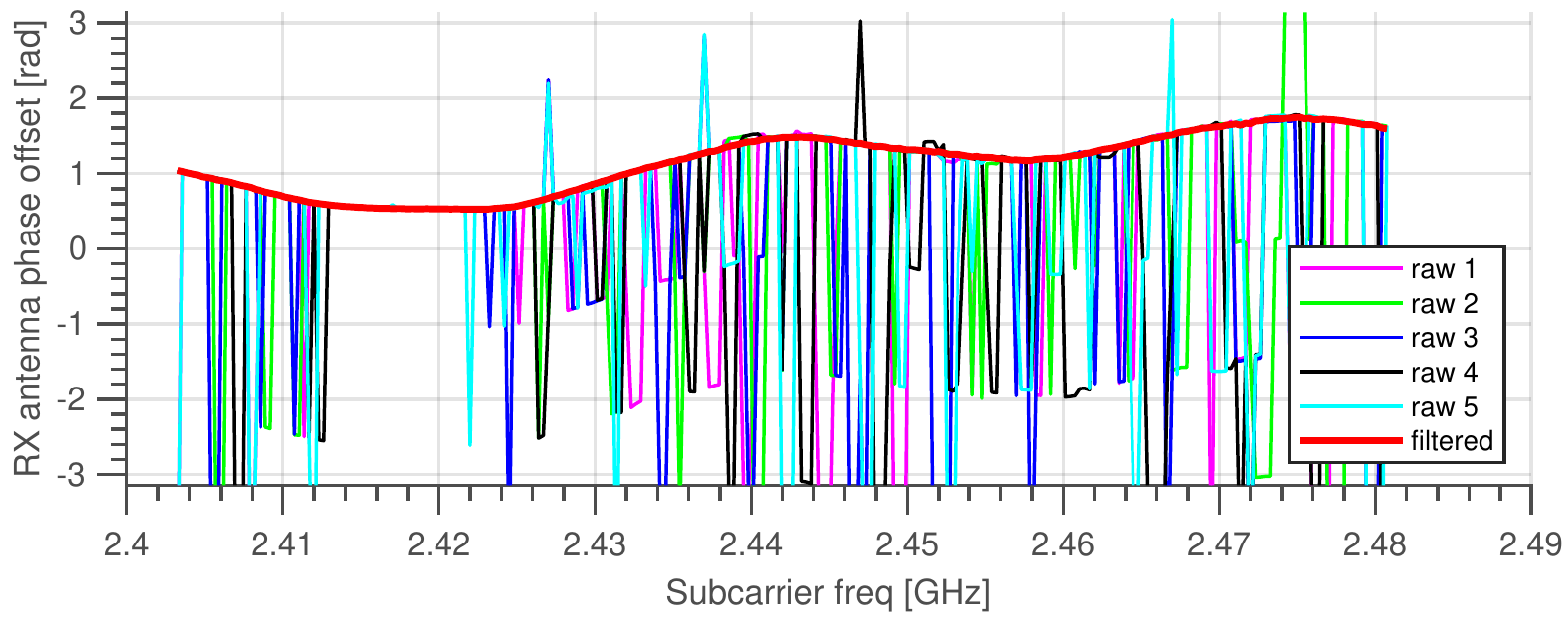}
    \vspace{-10pt}
    \caption{Example of $\phi$ derived directly from raw data (1-5) vs. after correction (red).}
    \label{fig:filter_ex}
    \vspace{-5pt}
\end{figure}

Fig.~\ref{fig:filter_ex} shows an example with $\hat{\phi}$ estimated directly from raw captured CSI and the final result after proposed correction method (red curve).
Note, that the filtering was performed over $13 \times 20 = 260$ packets, i.e. 20 packets transmitted on each of the 13 channels.
Note that although the proposed approach requires the transmission of a large number of packets it is still useful as it does not require and additional calibration procedure.

%

\section{Case Study - AoA}
As proof-of-concept and a way to verify our results we run over-the-air experiments to estimate the Angle of Arrival (AoA) of the transmitter.

\subsection{Experiment}
\begin{figure}[ht]
    \centering
    \includegraphics[width=0.65\linewidth]{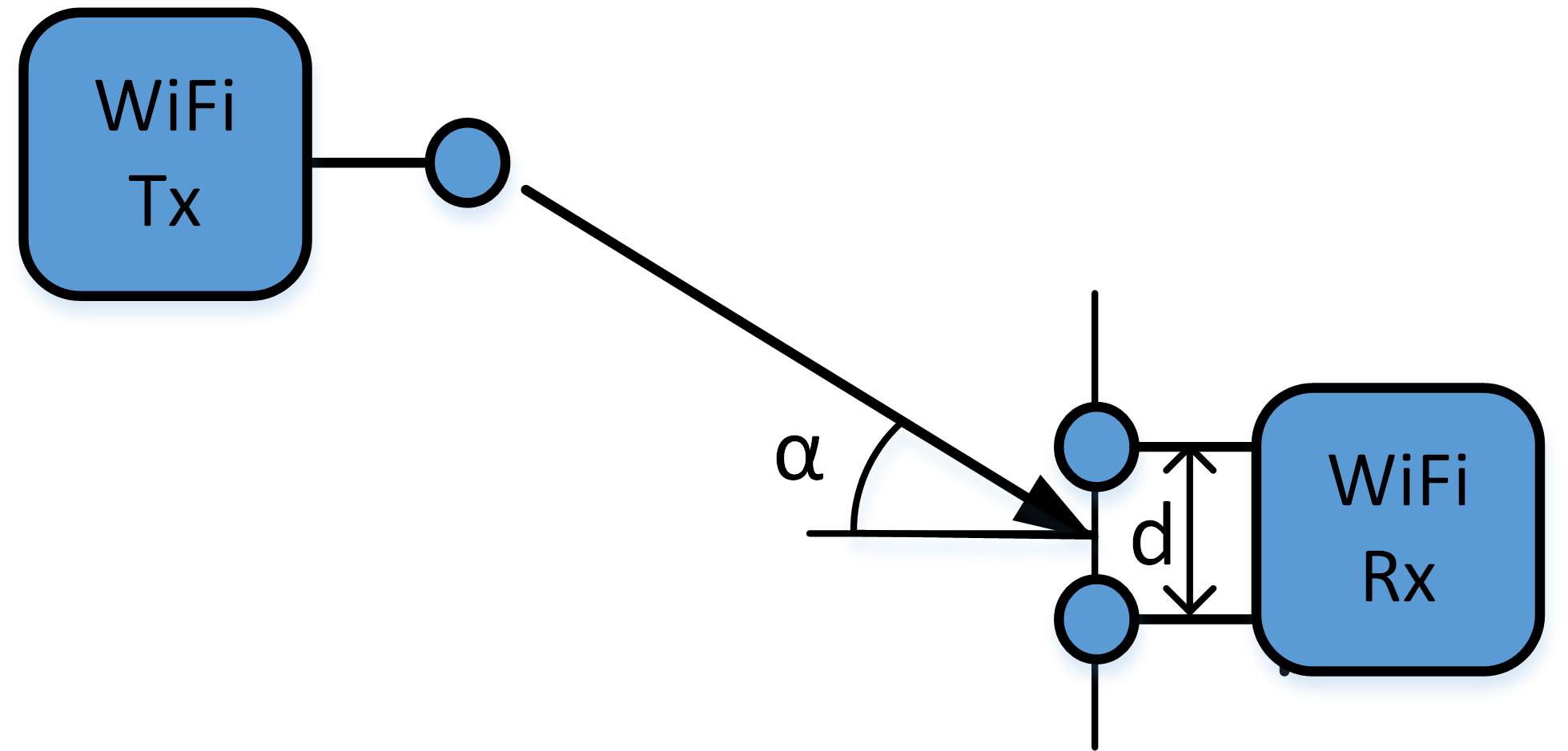}
    \caption{AoA experiment setup: transmitter placed at different angle $\alpha$ to receiver.}
    \label{fig:aoa_exp_setup}
    \vspace{-5pt}
\end{figure}
The experiment setup consists of a transmitter node with single antenna used, whereas the receiver node had two antennas being spaced by d=9\,cm (Fig.~\ref{fig:aoa_exp_setup}).
We analyzed seven different transmitter locations with ground truth AoA $\alpha \in \{-23\degree,-16\degree,-9\degree,-2\degree,5\degree,12\degree,19\degree\}$ while keeping the distance between the two nodes the same.
In each location, we sent 10k packets (HT20, MCS 0, BPSK) on all 13 channels in 2.4\,GHz band. 
%
%
Note, in total we had 12 channel switches.
From each received packet we collected the CSI and annotated with the channel used.
During post-processing in Matlab we created a single 80\,MHz channel by stitching together all channels.
Moreover, we corrected the estimated RX phase offset using the approach from \S\ref{sec:cleansing} and also corrected the fixed offset between antennas as explained in \S\ref{sec:results}.
Finally, similar to \cite{zhang2019calibrating} we estimated the AoA using algorithm as proposed in \cite{kotaru2015spotfi}.

\subsection{Results}
\begin{figure}[ht]
    \centering
    \includegraphics[width=1\linewidth]{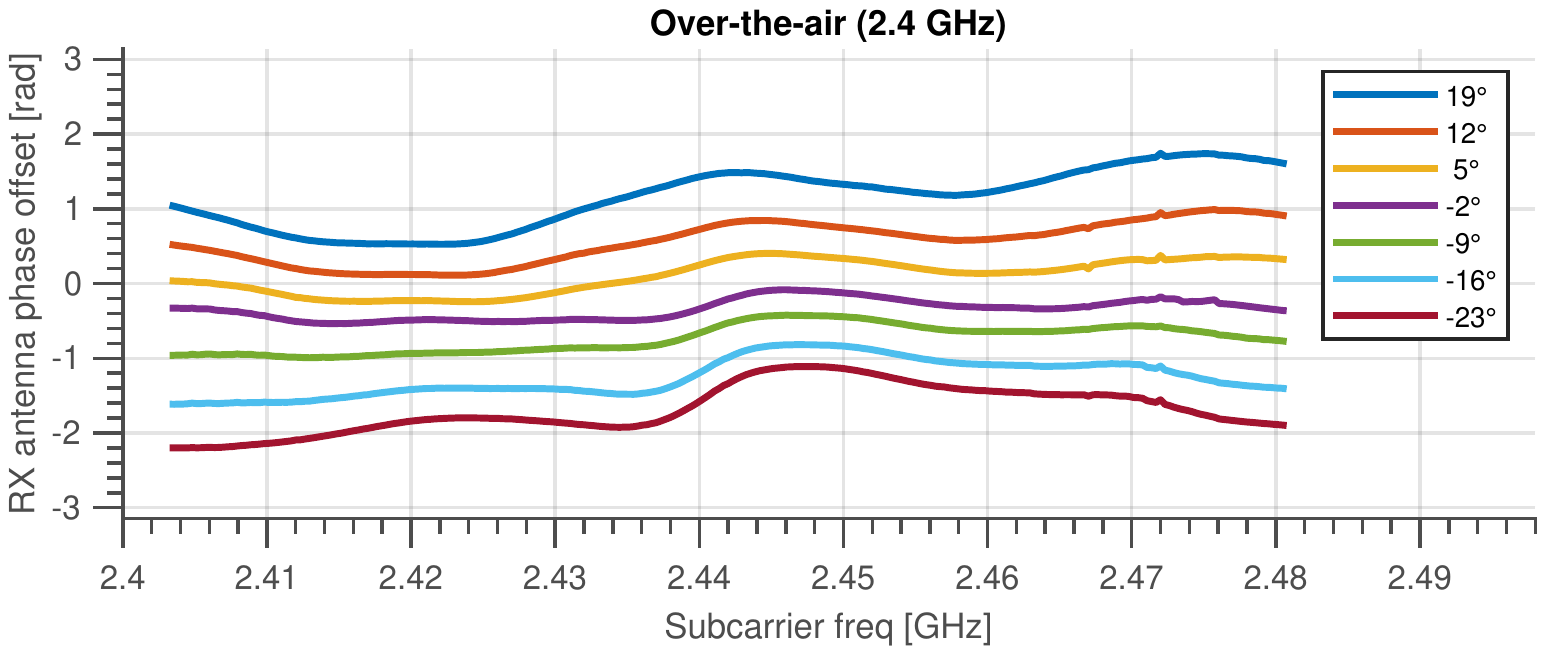}
    \vspace{-10pt}
    \caption{Estimated phase offset between the two receive antennas for the six different AoA values.}
    \label{fig:air_rxphaseoffset_24ghz}
    \vspace{-5pt}
\end{figure}

\begin{figure}[ht]

    \centering
    \includegraphics[width=1\linewidth]{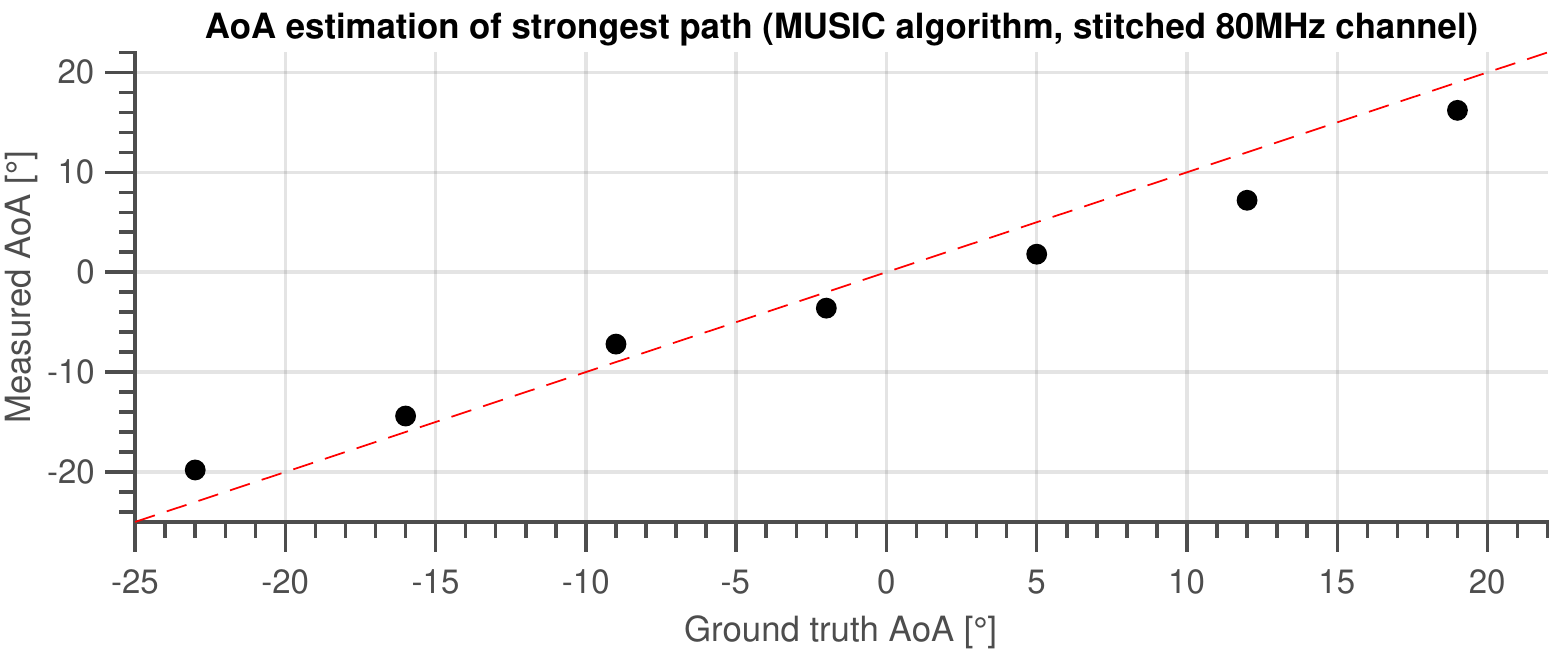}
    \vspace{-10pt}
    \caption{AoA estimated with algorithm \cite{kotaru2015spotfi} vs. ground truth.}
    \label{fig:air_exp_music_algo}
    \vspace{-5pt}
\end{figure}

Fig.~\ref{fig:air_rxphaseoffset_24ghz} shows the estimated phase offset between the two receive antennas for all subcarriers in the combined 80\,MHz channel for the seven transmitter locations.
As expected, the RX phase offset changes smoothly from one subcarrier to another.
Finally, Fig.~\ref{fig:air_exp_music_algo} shows the estimated AoA vs. the ground truth.
The former was obtained from 260 packets sent on the 13 channels in 2.4\,GHz band.
We can observe shows very good accuracy.

%

\section{Conclusions}
In this paper we analyzed the phase offset between receive antennas of modern 802.11ac NICs using Intel 9260 chips.
We can confirm that the phase offset between receive antennas is due to random phase rotations semi-time-invariant with up to four possible values.
Moreover, it is frequency dependent, i.e. subcarrier used.
Therefore, we presented an algorithm for cleansing the CSI to derive the true receive phase offset.
As proof-of-concept and a way to verify our results, we implemented an Angle of Arrival (AoA) algorithm, which showed very good performance.

As future work, we plan to extend your analysis towards using wider channels, i.e. 40, 80 and 160\,MHz, in 802.11ac.

\bibliographystyle{IEEEtran}
\bibliography{biblio,IEEEabrv}

\begin{thebibliography}{1}
\providecommand{\url}[1]{#1}
\csname url@samestyle\endcsname
\providecommand{\newblock}{\relax}
\providecommand{\bibinfo}[2]{#2}
\providecommand{\BIBentrySTDinterwordspacing}{\spaceskip=0pt\relax}
\providecommand{\BIBentryALTinterwordstretchfactor}{4}
\providecommand{\BIBentryALTinterwordspacing}{\spaceskip=\fontdimen2\font plus
\BIBentryALTinterwordstretchfactor\fontdimen3\font minus
  \fontdimen4\font\relax}
\providecommand{\BIBforeignlanguage}[2]{{%
\expandafter\ifx\csname l@#1\endcsname\relax
\typeout{** WARNING: IEEEtran.bst: No hyphenation pattern has been}%
\typeout{** loaded for the language `#1'. Using the pattern for}%
\typeout{** the default language instead.}%
\else
\language=\csname l@#1\endcsname
\fi
#2}}
\providecommand{\BIBdecl}{\relax}
\BIBdecl

\bibitem{ma2019wifi}
Y.~Ma, G.~Zhou, and S.~Wang, ``Wifi sensing with channel state information: A
  survey,'' \emph{ACM Computing Surveys (CSUR)}, vol.~52, no.~3, pp. 1--36,
  2019.

\bibitem{kotaru2015spotfi}
M.~Kotaru, K.~Joshi, D.~Bharadia, and S.~Katti, ``Spotfi: Decimeter level
  localization using wifi,'' in \emph{Proceedings of the 2015 ACM Conference on
  Special Interest Group on Data Communication}, 2015, pp. 269--282.

\bibitem{zhang2019calibrating}
D.~Zhang, Y.~Hu, Y.~Chen, and B.~Zeng, ``Calibrating phase offsets for
  commodity wifi,'' \emph{IEEE Systems Journal}, 2019.

\bibitem{zhuo2017perceiving}
Y.~Zhuo, H.~Zhu, H.~Xue, and S.~Chang, ``Perceiving accurate csi phases with
  commodity wifi devices,'' in \emph{IEEE INFOCOM 2017-IEEE Conference on
  Computer Communications}.\hskip 1em plus 0.5em minus 0.4em\relax IEEE, 2017,
  pp. 1--9.

\bibitem{gawlowicz2017uniflex}
P.~Gawłowicz, A.~Zubow, M.~Chwalisz, and A.~Wolisz, ``{UniFlex: A Framework
  for Simplifying Wireless Network Control},'' in \emph{IEEE International
  Conference on Communications (ICC 2017)}.\hskip 1em plus 0.5em minus
  0.4em\relax Paris, France: Institute of Electrical and Electronics Engineers,
  5 2017.

\bibitem{halperin2011tool}
D.~Halperin, W.~Hu, A.~Sheth, and D.~Wetherall, ``Tool release: Gathering
  802.11 n traces with channel state information,'' \emph{ACM SIGCOMM Computer
  Communication Review}, vol.~41, no.~1, pp. 53--53, 2011.

\bibitem{xiong2013arraytrack}
J.~Xiong and K.~Jamieson, ``Arraytrack: A fine-grained indoor location
  system,'' in \emph{Presented as part of the 10th $\{$USENIX$\}$ Symposium on
  Networked Systems Design and Implementation ($\{$NSDI$\}$ 13)}, 2013, pp.
  71--84.

\end{thebibliography}

\end{document}